\documentclass[12pt]{article}
\usepackage{latexsym}
\usepackage{amsmath}
\usepackage{amssymb}
\usepackage{verbatim}
\usepackage[english,french]{babel}
\usepackage[T1]{fontenc}
\newcommand{\xp}[1]{$^{\mbox{\scriptsize #1}}$}
\newcommand{\bibspace}{\vspace{1ex}\noindent}
\begin{document}
\begin{center}
\Large
\textbf{La réalité face à la
théorie quantique}\\[0.5cm]
\large
\textbf{Louis Marchildon}\\[0.5cm]
\normalsize
D\'{e}partement de chimie, biochimie et physique,\\
Universit\'{e} du Qu\'{e}bec,
Trois-Rivi\`{e}res, Qc.\ Canada G9A 5H7\\
\verb+louis.marchildon@uqtr.ca+
\end{center}
%
%
\begin{abstract}
Tous les chercheurs intéressés aux fondements
de la théorie quantique s'entendent sur le fait
que celle-ci a profondément modifié notre
conception de la réalité. Là s'arrête, toutefois,
le consensus. Le formalisme de la théorie,
non problématique, donne lieu à plusieurs
interprétations très différentes, qui ont
chacune des conséquences sur la notion de
réalité. Cet article analyse comment
l'interprétation de Copenhague, l'effondrement
du vecteur d'état de von Neumann, l'onde pilote
de Bohm et de Broglie et les mondes multiples
d'Everett modifient, chacun à sa manière,
la conception classique de la réalité, dont
le caractère local, en particulier, requiert
une révision.

\selectlanguage{english}
\begin{center}
\textbf{Abstract}
\end{center}

\vspace{-1ex}
All investigators working on the foundations
of quantum mechanics agree that the theory
has profoundly modified our conception of
reality. But there ends the consensus. The
unproblematic formalism of the theory
gives rise to a number of very different
interpretations, each of which has consequences
on the notion of reality. This paper analyses
how the Copenhagen interpretation, von Neumann's
state vector collapse, Bohm and de Broglie's
pilot wave and Everett's many worlds modify,
each in its own way, the classical conception
of reality, whose local character, in particular,
requires revision.
\end{abstract}
%
\selectlanguage{french}
\section{Introduction}
De toutes les révolutions qui ont marqué
le progrès de la science, aucune n'a changé
notre conception de la réalité autant que
l'avènement de la théorie quantique.
Certes, d'autres bouleversements scientifiques
ont profondément modifié le rapport
intellectuel que leurs contemporains
entretenaient avec le monde dans lequel
ils vivaient.  Les humains qui, avant Nicolas
Copernic, se croyaient situés au centre de
l'univers sur une Terre immobile, ont
progressivement compris que le mouvement
de celle-ci les emportait à grande vitesse
autour du Soleil.  La théorie de la
sélection naturelle de Charles Darwin a
sonné le glas des explications finalistes
auxquelles philosophes et scientifiques
recouraient depuis Platon et Aristote.
Et la théorie de la relativité restreinte
d'Albert Einstein a soustrait à l'espace
et au temps le caractère absolu qu'Isaac
Newton leur avait accordé.  Avec la théorie
quantique, toutefois, l'individualité des
objets, ou la notion même de matière,
semblent se dérober.

La théorie quantique a renversé ce qu'on
appelle la conception \emph{classique} de
la réalité.  Ce terme recouvre lui-même
plusieurs notions, et la théorie classique
a beaucoup évolué entre Galilée, au début du
\textsc{xvii}\xp{e} siècle, et Max Planck
et Einstein, au début du \textsc{xx}\xp{e}.
Néanmoins, on peut
soutenir que les deux concepts les plus
fondamentaux de la physique classique sont
celui de particule et celui de champ.

Le concept de particule remonte aux
atomistes grecs, mais il est sorti transformé
de l'analyse mathématique de Newton. Une
particule est un objet qui a une masse,
qui occupe à tout instant une position,
qui se déplace à une vitesse bien définie
et qui est soumis à des forces. Les
trajectoires des particules (c'est-à-dire
leurs positions comme fonctions du temps)
obéissent à des équations différentielles
qui, lorsqu'on spécifie des conditions
initiales appropriées, ont des solutions
uniques. Pierre Simon de Laplace conclura
que l'évolution du système du monde est
parfaitement déterminée.

La notion de champ, introduite au milieu
du \textsc{xix}\xp{e} siècle par Michael
Faraday et James Clerk Maxwell, paraît d'abord
compliquer la perspective. À l'origine,
le champ sert à représenter l'effet
d'une charge ou d'un courant électrique
sur une autre charge ou un autre courant,
sans faire appel à l'action à distance
que la gravité newtonienne semblait
exiger. Mais le champ est conçu, à l'origine,
comme une déformation d'un milieu matériel
qu'on appelle \emph{l'éther}, à la fois
extrêmement ténu et très rigide. La
théorie de la relativité restreinte
d'Einstein permettra d'éliminer l'éther,
à partir du point de vue où les champs se
propagent dans le vide. Il reste que le champ,
qui possède énergie, impulsion et moment
cinétique, est bien réel.

Les deux forces fondamentales de la
physique classique, la gravitation et
l'électromagnétisme, ont en principe une portée
infinie. Étant donné, toutefois, qu'elles
diminuent en raison du carré de la distance,
leur effet loin de la source devient à toutes
fins pratiques négligeable. Ainsi, un ensemble
d'objets rapprochés les uns des autres, mais très
éloignés de tous les autres objets (par exemple,
les corps du système solaire), peut être
considéré, sans trop d'erreur, comme un
système isolé. Cette propriété simplifie
singulièrement l'analyse conceptuelle des
systèmes classiques.

Les lois fondamentales de la physique classique
s'énoncent typiquement au moyen d'équations
dont les variables s'interprètent directement
en termes de particules et de champs. En
théorie quantique, par contre, l'interprétation
des objets du formalisme (vecteur d'état,
opérateurs, comme nous le verrons) est beaucoup
moins directe. À cela s'ajoutent deux problèmes
qui, absents de la physique classique, ont en
théorie quantique de profondes conséquences sur
la nature de la réalité: il s'agit de la mesure
et des corrélations à distance.

Après un bref survol du formalisme de la
théorie quantique, nous allons présenter le
problème de la mesure à la section~2. Nous
introduirons ensuite, aux sections~3 à~6,
quatre interprétations distinctes de la
théorie quantique et verrons comment chacune
tente de le résoudre. À la section~7, nous
examinerons les corrélations à distance et
signalerons comment chaque interprétation
essaie d'en rendre compte. Nous conclurons
à la section~8.

Au cours de sa longue carrière, Mario Bunge
a résolument défendu une approche réaliste
de la philosophie des sciences, qu'il a
appelée le \emph{réalisme critique}. Le réalisme
suscite, en ce début du \textsc{xxi}\xp{e}
siècle, un intérêt renouvelé. Différentes
interprétations de la théorie quantique
donnent lieu à différentes conceptions de
la réalité, que nous essaierons de mettre en
lumière sans toutefois prendre position pour
l'une ou pour l'autre. Nous aurons l'occasion
de signaler plusieurs apports de Bunge à la
réflexion sur la théorie quantique, qu'il a
présentés entre autres dans les ouvrages cités
dans la bibliographie.
%
\section{Le problème de la mesure}
Par \emph{grandeur physique}, nous entendons
toute quantité susceptible de mesure. En
théorie classique, la position d'une
particule, son énergie cinétique, la valeur
du champ électrique en un point, la pression
et la température d'un gaz sont, pour n'en
nommer que quelques-unes, des grandeurs
physiques. On admet (toujours en théorie
classique) que toutes les grandeurs physiques
ont, à tout instant, des valeurs bien définies.

Pour mesurer une grandeur associée à
un système physique, on utilise un appareil.
Celui-ci comporte un \emph{indicateur},
c'est-à-dire un dispositif qui peut
prendre différentes valeurs. Le processus
de mesure consiste en une interaction entre
l'appareil et le système. Pour que la mesure
soit fidèle, il faut que l'indicateur marque,
au terme de l'interaction, une valeur qui
caractérise la valeur initiale de la
grandeur physique.

À titre d'exemple, on peut mesurer
la température d'un liquide en y plongeant
un thermomètre, où la colonne de mercure
indique la température ambiante. Strictement
parlant, l'interaction a pour effet
de perturber le système physique. Mais
en théorie classique, on peut toujours,
en principe, réduire cette perturbation
à un niveau arbitrairement faible. Il n'y
a aucune limite, par exemple, à la réduction
de la dimension du thermomètre utilisé.

En théorie quantique, on ne peut réduire
indéfiniment l'interaction entre le système
et l'appareil. Qui plus est, le processus
de mesure présente un trait problématique
qu'on ne retrouve d'aucune manière en
théorie classique. Avant de l'analyser,
cependant, nous allons brièvement
rappeler quelques notions du formalisme que
nous devrons utiliser.\footnote{Le lecteur
familier du formalisme, ou celui qui
souhaite l'éviter le plus possible,
peuvent se reporter à l'éq.~(\ref{vn1}).
On trouvera un exposé plus complet dans les
traités de base de mécanique quantique. Nous
utilisons la notation de Marchildon (2000).}
\subsubsection*{Espace d'états}
À tout système quantique est associé un
espace vectoriel complexe (de dimension finie
ou infinie) qu'on peut noter~$\mathcal{V}$
et qu'on appelle \emph{l'espace d'états}.
Cet espace est muni d'un produit scalaire,
qui permet de définir la \emph{norme} d'un
vecteur. Tout vecteur non nul~$|\psi\rangle$
dans~$\mathcal{V}$ représente un état possible
du système (on désigne alors~$|\psi\rangle$
sous le nom de \emph{vecteur d'état}). La
correspondance, toutefois, n'est pas biunivoque,
puisque deux vecteurs multiples l'un de l'autre
représentent le même état. On peut donc
toujours remplacer un vecteur d'état par un
vecteur équivalent de norme~1. Si~$|\psi_1\rangle$
et~$|\psi_2\rangle$ représentent deux états
distincts d'un système quantique et si~$c_1$
et~$c_2$ sont deux nombres complexes, alors le
vecteur
\begin{equation}
|\psi\rangle = c_1 |\psi_1\rangle
+ c_2 |\psi_2\rangle
\label{comb}
\end{equation}
appartient à~$\mathcal{V}$ et représente lui
aussi un état possible du système. C'est le
\emph{principe de superposition}.

Dans plusieurs situations, on ne peut
représenter l'état d'un système par un simple
vecteur~$|\psi\rangle$. On doit alors utiliser
une distribution pondérée de vecteurs ou, ce
qui revient au même, un \emph{opérateur
densité}. Il ne faut pas confondre une
distribution pondérée de vecteurs avec une
combinaison linéaire comme~(\ref{comb}).
\subsubsection*{Grandeur physique}
Un \emph{opérateur} est un objet mathématique
qui transforme un vecteur en un autre vecteur.
Dans le cas où un opérateur~$O$ transforme
un vecteur~$|\phi\rangle$ par un facteur
multiplicatif~$\lambda$ (c'est-à-dire que
$O |\phi\rangle = \lambda |\phi\rangle$),
on dit que~$|\phi\rangle$ est un \emph{vecteur
propre} de~$O$ correspondant à la \emph{valeur
propre}~$\lambda$.
  
En théorie quantique, on associe habituellement
à une grandeur physique un opérateur
\emph{auto-adjoint}. Un opérateur
auto-adjoint est linéaire (son action
sur une somme ou une différence de vecteurs
est la somme ou la différence des actions),
et toutes ses valeurs propres sont des
nombres réels.

Soit~$A$ l'opérateur auto-adjoint associé
à une grandeur physique (que nous noterons
également~$A$). En théorie quantique, on admet
que les seuls résultats possibles de la mesure
de~$A$ sont ses valeurs propres (notées~$a_1$,
$a_2$, \ldots).\footnote{On parle ici, bien
sûr, d'une mesure idéale, puisque les mesures
concrètes ont toujours une précision finie.
Au delà d'un point de vue opérationnel, on
peut chercher à associer les valeurs propres non
pas simplement aux résultats de mesure mais,
de façon réaliste, aux seules valeurs possibles
de la grandeur physique (Bunge, 1967). Il s'agit
toutefois d'une hypothèse supplémentaire, que
plusieurs interprétations de la théorie ne
retiennent pas.} Soit~$|a_i\rangle$ un vecteur
propre (de norme~1) correspondant à la valeur
propre~$a_i$. Si, à un instant donné, le vecteur
d'état du système quantique coïncide
avec~$|a_i\rangle$, alors la mesure de la
grandeur~$A$ donnera nécessairement la
valeur~$a_i$.
\subsubsection*{Règle de Born}
Supposons que toutes les valeurs propres
de~$A$ soient discrètes et
distinctes.\footnote{L'extension au cas où
certaines valeurs propres coïncident est
aisée, mais nous n'en aurons pas besoin.}
Supposons de plus qu'à un instant donné,
le système quantique se trouve dans l'état
représenté par le vecteur~$|\psi\rangle$. On
peut montrer qu'il est possible
d'exprimer~$|\psi\rangle$ comme une
combinaison linéaire des vecteurs propres
de~$A$, c'est-à-dire
\begin{equation}
|\psi\rangle = \sum_i c_i |a_i\rangle ,
\label{combin}
\end{equation}
où les $c_i$ sont des nombres complexes.

La \emph{règle de Born} énonce que si l'on
mesure la grandeur~$A$ sur le système dans
l'état~$|\psi\rangle$, la probabilité
d'obtenir le résultat~$a_i$ est égale
à~$|c_i|^2$.\footnote{Ici encore, on peut
chercher à énoncer la règle de Born de façon
réaliste en associant à~$|c_i|^2$ la
\emph{propension} de la grandeur~$A$ à avoir
la valeur~$a_i$ (Popper, 1982; Bunge, 1985).
On ne peut pas, cependant, associer aux
probabilités~$|c_i|^2$ une ignorance de
valeurs précises que toutes les grandeurs
physiques auraient (Kochen et Specker, 1967).}
\subsubsection*{Évolution temporelle}
Jusqu'à maintenant, nous avons toujours
considéré le vecteur d'état à un instant
précis. Le fait est, cependant, que le
vecteur d'état d'un système quantique (comme
la position d'une particule classique) change
avec le temps. On le note donc de
façon plus complète comme~$|\psi (t)\rangle$.
Considérons un système quantique isolé,
et soit~$|\psi (t_0)\rangle$ le vecteur
d'état à l'instant initial~$t_0$. Alors le
vecteur d'état à tout instant~$t$ est
donné par
\begin{equation}
|\psi (t)\rangle = U(t, t_0) |\psi (t_0)\rangle .
\label{evol}
\end{equation}
Ici, $U$ est un opérateur \emph{unitaire},
c'est-à-dire un opérateur linéaire qui ne
change pas la norme du vecteur sur lequel
il agit. L'opérateur~$U$ dépend de la
dynamique du système. L'éq.~(\ref{evol})
est équivalente à une équation différentielle
qu'on appelle \emph{l'équation de Schrödinger}.

\bigskip
Nous pouvons maintenant revenir à la mesure.
Notons d'abord que, concrètement, un appareil
consiste en un objet macroscopique, composé
d'un nombre d'atomes immense, de l'ordre du
nombre d'Avogadro. Dans une perspective
réductionniste, on peut penser que la théorie
quantique, qui régit le comportement de chaque
atome, régit également l'agrégat d'atomes qui
constitue l'appareil. Le processus de mesure
d'une grandeur d'un système quantique par un
appareil devrait donc, en principe, pouvoir
être décrit entièrement par le formalisme de
la théorie quantique. À juste titre, Bunge a
signalé qu'on ne peut donner une théorie
détaillée de la mesure qui s'applique en toute
généralité, puisque la mesure de grandeurs
distinctes exige l'utilisation d'appareils
complètement différents. Néanmoins, sans devoir
recourir à la description détaillée de
dispositifs spécifiques, nous allons voir que
la seule hypothèse du comportement quantique de
l'appareil conduit à un problème fondamental.

Soit~$\mathcal{S}$ un système quantique dont
on cherche à mesurer une grandeur~$A$ et
soit~$\mathcal{M}$ l'appareil dont on
dispose pour ce faire. Bien que très complexe,
$\mathcal{M}$ est lui aussi un système quantique.
On suppose que le système~$\mathcal{S}$ et
l'appareil~$\mathcal{M}$ constituent un
grand système quantique isolé
$\mathcal{S} \oplus \mathcal{M}$. La mesure
de la grandeur~$A$ consistera en un processus
dynamique à travers lequel ce grand système
évolue.

Supposons qu'au départ, l'état de l'appareil
soit représenté par le vecteur~$|\alpha_0\rangle$.
Pour que le processus de mesure soit fidèle,
il faut que si la grandeur~$A$ possède
initialement une valeur bien définie, l'appareil
indique cette valeur au terme de la mesure.
Cela implique qu'il doit y avoir dans
l'espace d'états de l'appareil des
vecteurs~$|\alpha_i\rangle$ qui représentent
des états de l'indicateur. La mesure sera donc
fidèle si, à travers l'interaction du système
avec l'appareil, l'état initial
$|a_i\rangle |\alpha_0\rangle$ (c'est-à-dire
$|a_i\rangle$ pour~$\mathcal{S}$ et
$|\alpha_0\rangle$ pour~$\mathcal{M}$) 
évolue vers l'état final
$|a_i\rangle |\alpha_i\rangle$
(c'est-à-dire $|a_i\rangle$ pour~$\mathcal{S}$ et
$|\alpha_i\rangle$
pour~$\mathcal{M}$).\footnote{Nous supposons,
pour plus de simplicité, que l'état
de~$\mathcal{S}$ n'est pas altéré par la mesure.}
On peut montrer que cette évolution s'effectue
au moyen d'un opérateur unitaire
$U(\mathcal{S}, \mathcal{M})$,\footnote{Nous
n'indiquons pas explicitement l'instant
initial et l'instant final du processus.}
agissant dans l'espace d'états du système et de
l'appareil. Symboliquement,
\begin{equation}
|a_i\rangle |\alpha_0\rangle \;
\raisebox{1.5ex}{$\underrightarrow{\;
\scriptstyle{U(\mathcal{S}, \mathcal{M})} \;\;}$}
\; |a_i\rangle |\alpha_i\rangle . 
\end{equation}
Il s'agit d'une condition nécessaire à
laquelle l'évolution doit satisfaire si
le processus de mesure est décrit par la
théorie quantique.

À titre d'exemple, le moment magnétique (ou
l'aimantation) d'un électron est une grandeur
physique dont la mesure peut donner deux valeurs
distinctes (disons~$+$ et~$-$). On mesure le
moment magnétique en dirigeant l'électron
vers un champ magnétique non homogène et en
notant la déviation de l'électron (la valeur~$+$
correspondant par exemple à une déviation vers
le haut et la valeur~$-$ à une déviation vers
le bas). Si l'électron est initialement dans
l'état~$|+\rangle$, il se dirigera nécessairement
vers le haut, et il se dirigera vers le bas s'il
est initialement dans l'état~$|-\rangle$. Les
vecteurs~$|+\rangle$ et~$|-\rangle$ correspondent
aux vecteurs~$|a_i\rangle$ du paragraphe
précédent.

Supposons maintenant que l'état initial
du système soit représenté par un vecteur
quelconque~$|\psi\rangle$.\footnote{Dans notre
exemple, $|\psi\rangle$ est une combinaison
linéaire $c_+ |+\rangle + c_- |-\rangle$.}
Utilisant l'éq.~(\ref{combin}), on voit que l'état
initial du système et de l'appareil est représenté
par
\begin{equation}
|\psi\rangle |\alpha_0\rangle
= \sum_i c_i |a_i\rangle |\alpha_0 \rangle .
\label{vn0}
\end{equation}
Étant donné la linéarité de l'opérateur~$U$,
on trouve que l'évolution du système et de
l'appareil est donnée par
\begin{equation}
\sum_i c_i |a_i\rangle |\alpha_0\rangle \;
\raisebox{1.5ex}{$\underrightarrow{\;
\scriptstyle{U(\mathcal{S}, \mathcal{M})} \;\;}$}
\; \sum_i c_i |a_i\rangle |\alpha_i\rangle . 
\label{vn1}
\end{equation}
Ainsi, dans l'état final, les valeurs finales de
l'indicateur sont parfaitement corrélées avec
les valeurs initiales de la grandeur physique.

Malheureusement, cette corrélation ne suffit
pas. En effet, rien n'indique dans
l'éq.~(\ref{vn1}) qu'un résultat
spécifique a été obtenu. Au contraire, la
superposition semble montrer que l'indicateur
marque en même temps tous les résultats
possibles. Ceci constitue, \emph{prima facie},
une violation flagrante de l'expérience.
Rappelons-nous, en effet, que l'appareil
est un système macroscopique. Jamais nous
ne voyons l'indicateur d'un appareil (qu'il
soit analogique ou numérique) marquer en
même temps deux résultats. L'incapacité
de la théorie quantique, comme nous l'avons
formulée, de rendre compte de l'obtention
d'un résultat spécifique s'appelle
\emph{le problème de la mesure}.

Le n{\oe}ud du problème de la mesure se situe
dans le caractère unitaire (et donc linéaire)
de l'évolution temporelle, et dans la
corrélation (ou \emph{l'intrication}) des
états du système et de l'appareil produite
par la dynamique. On peut le formuler sans
référence directe à la notion d'appareil,
comme Erwin Schrödinger l'a fait au moyen du
chat qu'on lui associe.\footnote{Schrödinger
(1935).} Schrödinger suppose qu'on isole
un chat dans une enceinte où l'on a préalablement
installé le dispositif suivant: un compteur
Geiger contient un atome radioactif d'un isotope
dont la demi-vie est égale à une heure; dès que
le compteur détecte la désintégration, il
déclenche un mécanisme qui libère dans
l'enceinte un gaz délétère, empoisonnant
aussitôt le chat. Dans ce cas, la vie et la mort
du chat sont parfaitement corrélées à
l'intégrité ou la désintégration de l'atome.

Le problème se pose dès qu'on considère non
seulement l'atome, mais aussi le poison
et le chat comme des systèmes quantiques.
Dans ce cas, une heure après le début de
l'expérience, le vecteur d'état du système
global est la somme de deux termes: un
premier, dans lequel l'atome est intègre,
le poison confiné et le chat vivant; et un
second, dans lequel l'atome s'est désintégré,
le poison s'est répandu et le chat est mort.
Or, personne n'a jamais vu un chat qui est
en même temps mort et vivant.

Dans les sections qui suivent, nous
examinerons quatre façons qu'on a
proposées pour résoudre le problème de la
mesure. Chacune a de profondes conséquences
sur notre conception de la réalité.
%
\section{L'interprétation de Copenhague}
La théorie quantique a connu une
gestation d'un quart de siècle. Sa
naissance, par contre, a été rapide. En une
année à peine, Werner Heisenberg, Paul Dirac
et Schrödinger ont proposé des formalismes
en apparence différents, mais qu'on a
rapidement reconnus équivalents. Les
applications ont alors explosé. La théorie
a expliqué la structure des atomes d'abord,
puis celle des noyaux atomiques et des solides,
et enfin celle des particules fondamentales.
Presque cent ans après son avènement, aucune
expérience ne l'a prise en défaut.

Aussi vite a-t-on appris à utiliser le
formalisme, aussi durement s'est-on heurté
à son interprétation. Ainsi Schrödinger,
après avoir réalisé que le vecteur d'état
d'un électron consiste en une fonction
complexe des coordonnées spatiales, a cru
pouvoir lier cette \emph{fonction d'onde}
à la distribution de charge de l'électron.
L'extension illimitée de la fonction l'a
convaincu de renoncer à l'idée. Max Born a
alors suggéré d'interpréter la fonction d'onde
(ou, plus exactement, son carré absolu) comme
la probabilité de trouver l'électron autour
de tel ou tel point. C'est l'origine de la
règle de Born que nous avons vue plus tôt.

L'interprétation du formalisme a constitué
un thème majeur du congrès Solvay de 1927,
qui a rassemblé tous les fondateurs de la
théorie quantique. Dans les années qui ont
suivi, Heisenberg et surtout Niels Bohr ont
élaboré une approche qu'on a appelée
\emph{l'interprétation de Copenhague}.
Bien que quelques fondateurs de la théorie
(Einstein, Schrödinger et Louis de Broglie)
ne s'y soient jamais ralliés, l'interprétation
de Copenhague est venu à dominer, pendant
quelques décennies, toute la réflexion sur
les fondements de la théorie quantique, au
point où l'historien Max Jammer a parlé de
{\og}la monocratie de l'école de
Copenhague{\fg}.\footnote{Jammer (1974),
p.~250.} Bunge a été l'un des premiers,
dans les années 1950, à s'y opposer et à
critiquer, en particulier, son caractère
subjectiviste.

D'emblée, l'interprétation de Copenhague adopte
la règle de Born et son caractère opérationnel.
On se rappelle que le formalisme de la théorie
quantique énonce que les seuls résultats
possibles de la mesure d'une grandeur physique
sont les valeurs propres de l'opérateur
associé. L'interprétation de Copenhague va
cependant plus loin. Elle affirme qu'on ne peut
attribuer de valeur précise à une grandeur
hors du cadre d'une mesure. Celle-ci, nous
l'avons vu, requiert l'utilisation d'un
appareil macroscopique. Bohr affirme qu'on doit
décrire le fonctionnement d'un tel appareil
par les lois de la physique classique. Il
est méthodologiquement incorrect de lui
attribuer un vecteur d'état. Ainsi, l'argument
qui conduit à la superposition~(\ref{vn1}) ne
peut véritablement décoller. Étant donné qu'un
objet classique se trouve toujours dans un
état bien défini, on ne peut d'aucune
manière le représenter par une superposition
d'états distincts.

L'interprétation de Copenhague attribue donc
différents niveaux de réalité à différents
systèmes: un appareil, régi par la théorie
classique, est bien réel, tandis que les
grandeurs physiques des systèmes quantiques
n'ont de réalité qu'au moment d'une mesure.
Bohr ne spécifie pas à quel point se situe
précisément la transition entre le quantique
et le classique: qu'est-ce qui fait qu'à un
certain niveau de complexité, un agrégat
d'atomes et de molécules devient un système
classique?

Au cours des dernières décennies, différents
chercheurs ont proposé des façons d'interpréter
la théorie quantique qui s'inspirent de
l'interprétation de Copenhague. Nous allons
en mentionner deux.

La première remonte à Heisenberg, et on la
désigne souvent par le terme \emph{épistémique}.
Ici, le vecteur d'état (malgré son nom) ne
représente pas l'état objectif d'un système
quantique, mais plutôt notre connaissance
de cet état. Dans cette veine se situe le
\emph{qubisme} (une abréviation de
\emph{Quantum Bayesianism}), élaboré surtout
par Christopher Fuchs et Rudiger Schack. Le
qubisme considère {\og}la mécanique quantique
[comme] un outil que quiconque peut utiliser
pour évaluer, sur la base de son expérience
passée, ses attentes probabilistes envers son
expérience subséquente{\fg}.\footnote{Fuchs
\emph{et al.} (2014). Ici comme ailleurs, la
traduction est nôtre.} Le qubisme adopte le point
de vue subjectif des probabilités, de sorte que
deux agents qui ont des expériences différentes
peuvent attribuer des probabilités différentes
à un même événement. Un agent peut utiliser la
théorie quantique pour modéliser n'importe
quel système extérieur à lui-même, y compris
d'autres agents. Selon le qubisme, {\og}la
mécanique quantique ne se rapporte pas directement
au monde objectif; elle se rapporte aux expériences
du monde objectif qui appartiennent à tout agent
particulier qui utilise la théorie
quantique{\fg}.\footnote{Fuchs \emph{et al.}
(2014).}

Le qubisme envisage le processus de mesure
de la manière suivante. Aussi longtemps que
l'agent n'a pas pris connaissance de la marque
de l'indicateur, l'état du système quantique
et de l'appareil est représenté par le membre
de droite de l'éq.~(\ref{vn1}). La superposition
reflète l'ignorance de l'agent. Par contre,
dès que celui-ci regarde l'indicateur, il
le voit marquer une valeur précise. Il
attribue donc comme vecteur d'état un seul
terme de la superposition. Cette transition
ne constitue pas un changement physique du
système et de l'appareil, mais la mise à jour
de l'expérience de l'agent. En définitive,
le qubisme résout le problème de la mesure
en refusant de considérer les propriétés
réelles des systèmes quantiques. Comme
d'autres approches instrumentalistes, il
exclut certains paramètres non observables
du champ de l'investigation
scientifique.\footnote{Marchildon (2015a).}

L'approche instrumentaliste la plus radicale
a été proposée par Ole Ulfbeck et Aage Bohr,
qui l'ont appelée \emph{genuine
fortuitousness}.\footnote{Ulfbeck et Bohr
(2001).} Pour ces
chercheurs, les particules quantiques
n'existent tout simplement pas. Un compteur
Geiger, rapproché d'un minerai d'uranium,
produit des clics selon une distribution
statistique bien définie. Mais il est incorrect, 
selon Ulfbeck et Bohr, d'attribuer ces clics
à la détection de particules alpha. Purement
fortuits, les clics n'ont aucune cause. La
théorie quantique vise à prédire la
probabilité de telle ou telle réaction d'un
appareil à une préparation expérimentale
donnée, l'appareil et la préparation étant
entièrement spécifiés en termes classiques.
Les objets macroscopiques sont bien réels,
mais il ne tirent pas ce trait de la réalité
d'hypothétiques constituants microscopiques.
%
\section{L'effondrement}
Le principe selon lequel le vecteur d'état
d'un système isolé évolue sous l'effet d'un
opérateur unitaire conduit à la superposition
représentée par le membre de droite de
l'éq.~(\ref{vn1}). John von Neumann a suggéré
que ce type d'évolution temporelle n'est pas
universel.\footnote{Von Neumann (1932).}

Von Neumann a soutenu que,
dans le contexte d'une mesure, il arrive un
moment où le membre de droite de~(\ref{vn1})
se réduit à un seul de ses termes. Lequel?
Il est impossible de le prévoir. On peut
toutefois établir des probabilités. Von
Neumann postule que la probabilité d'arriver
au terme $|a_i\rangle |\alpha_i\rangle$ est
égale à $|c_i|^2$, le carré absolu du coefficient
de ce terme dans la superposition~(\ref{vn1}).
Ainsi, il retrouve la règle de Born. Le passage
aléatoire de la superposition~(\ref{vn1}) à un
seul de ses termes s'appelle \emph{l'effondrement
du vecteur d'état}.

Examinons donc plus spécifiquement
l'effondrement. Rien n'oblige, en réalité, de
le situer au stade où nous l'avons introduit.
Manifestement, le chercheur qui réalise une
expérience de laboratoire observe que
l'indicateur de l'appareil marque un seul
résultat. Cherchons à décrire ce processus
d'observation par le formalisme de la
théorie quantique. Examinons, par exemple,
les photons qui tombent sur l'indicateur et
sont réfléchis dans l'{\oe}il du chercheur.
L'endroit où ces photons touchent la rétine
est alors corrélé avec la position de
l'indicateur, de sorte que le vecteur d'état
global (du système quantique, de l'indicateur
et de la rétine) fait intervenir une
superposition de tous ces endroits. Allons
plus loin. L'excitation de tels ou tels
neurones du cortex visuel est corrélée avec
l'endroit de la rétine où les photons sont
arrivés. Le vecteur d'état global (du système
quantique, de l'indicateur, de la rétine et
des neurones) fait donc intervenir une
superposition de ces excitations. D'un point
de vue pratique, peu importe à quel stade a lieu
l'effondrement du vecteur d'état. Von Neumann
remarque, cependant, qu'il doit avoir lieu
quelque part, puisque le chercheur a conscience
de n'observer qu'un seul résultat.

Quel impact cette analyse a-t-elle sur la notion
de réalité? En fait, plus l'effondrement
avance dans la chaîne causale, moins de
grandeurs physiques (la position de l'indicateur,
l'endroit où la rétine est stimulée, etc.)
n'ont de valeurs bien définies. Dans cette veine,
Fritz London, Edmond Bauer et, plus tard, Eugene
Wigner ont proposé une hypothèse
audacieuse:\footnote{London et Bauer (1939);
Wigner (1961).} ils ont suggéré que l'effondrement
du vecteur d'état requiert l'intervention d'un
sujet conscient. De ce point de vue, la
dualité esprit-matière réapparaît de façon
spectaculaire. Il faut dire que l'hypothèse
de London, Bauer et Wigner retient peu d'adeptes
aujourd'hui. À la section~6, cependant, nous
la verrons ressurgir sous une autre forme.

Von Neumann a introduit l'effondrement sous
la forme d'un postulat, sans suggérer de
mécanisme par lequel ce processus s'accomplirait.
Cela dit, pourquoi ne pas {\og}admettre que
l'effondrement se produit mais essayer de
\emph{réformer} cette hypothèse, la transformant
en un résultat approximatif dérivable des
principes physiques usuels appliqués à un
strict processus physique dans lequel des objets
quantiques interagissent avec des objets
classiques.{\fg}\footnote{Bunge (1985), p.~201.}
Gian Carlo Ghirardi, Alberto Rimini et Tullio
Weber ont justement proposé une théorie spécifique
de l'effondrement.\footnote{Ghirardi \emph{et al.}
(1986); voir aussi Ghirardi \emph{et al.} (1990).}

Ghirardi, Rimini et Weber (GRW) considèrent le
vecteur d'état d'un ensemble de $N$~particules,
qu'ils expriment sous la forme d'une fonction d'onde
$\psi(\mathbf{r}_1, \ldots, \mathbf{r}_N, t)$.
Le carré absolu de la fonction d'onde est
proportionnel à la probabilité conjointe
de trouver, à l'instant~$t$, la première
particule autour du point~$\mathbf{r}_1$, la
seconde autour du point~$\mathbf{r}_2$, etc.
GRW supposent que, la plupart du temps, la
fonction d'onde évolue sous l'effet d'un
opérateur unitaire~$U$ ou, ce qui revient
au même, par l'équation de Schrödinger.

Mais il y a plus. De temps à autre,
et de façon aléatoire, l'une ou l'autre
des particules subit une \emph{localisation
spontanée}, qui fait que la fonction d'onde
s'annule (à toutes fins pratiques) si les
coordonnées de la particule s'éloignent d'une
valeur donnée. La fréquence et la précision de
la localisation sont régies par deux paramètres
qui, selon GRW, constituent deux constantes
de la nature. S'il n'y a qu'une seule particule,
la fréquence de localisation est très faible.
Par contre, GRW montrent que la fréquence de
localisation du centre de masse d'un corps
rigide est égale à la somme des fréquences de
localisation de ses constituants. Ainsi, un
appareil macroscopique est constamment
localisé, ce qui constitue essentiellement
un effondrement du vecteur d'état.

La théorie de GRW résout le problème de la
mesure. Elle fait de plus, dans certaines
circonstances, des prédictions différentes de
celles de la théorie quantique proprement
dite. Pour l'instant, l'expérience n'a pas
réussi à les départager, mais il est bien
possible qu'on y parvienne dans un avenir
assez proche.

Quelle est, du point de vue de la théorie de
GRW, la nature de la réalité? Deux réponses
fort différentes ont été
proposées.\footnote{Allori \emph{et al.}
(2008).} La première associe à l'ensemble des
$N$~particules une densité de masse donnée par
\begin{equation}
m(\mathbf{r}, t) = \sum_{i=1}^N m_i
\int d\mathbf{r}_1 \ldots d\mathbf{r}_N
\delta(\mathbf{r}_i - \mathbf{r})
|\psi(\mathbf{r}_1, \ldots, \mathbf{r}_N, t)|^2 .
\label{masse}
\end{equation}
Ici, $m_i$ est la masse de la particule~$i$.
L'éq.~(\ref{masse}) représente une somme
pondérée (par les masses~$m_i$) de la densité
de probabilité de la position de chaque
particule. Cette association s'apparente à
l'idée originale de Schrödinger à propos de
la densité de charge électronique. Néanmoins,
elle ne souffre pas, pour les objets
macroscopiques, de l'objection liée à la
dispersion rapide de la densité puisque,
dans la théorie de~GRW, celle-ci est
constamment localisée. À toutes fins pratiques,
on ne peut distinguer, pour un objet
macroscopique, la densité de masse~(\ref{masse})
d'une densité classique.

La seconde façon possible d'envisager la
réalité est beaucoup moins intuitive. Il
s'agit de ne considérer comme réels que les
points spatiotemporels de localisation.
Dans tout volume spatial et dans tout intervalle
de temps finis, il n'y a qu'un nombre fini de
tels points. Certes, ce nombre est très grand
pour un objet macroscopique. Quoi qu'il en soit,
cette façon d'envisager la réalité diffère
singulièrement de la vision d'un ensemble de
particules qui se déplacent selon des
trajectoires bien définies, point de vue vers
lequel nous allons maintenant nous tourner.
%
\section{L'onde pilote}
L'interprétation de Copenhague et la théorie
de l'effondrement constituent deux manières
différentes de comprendre la théorie
quantique. Elles se rejoignent cependant
sur un point: dans les deux approches, le
comportement d'un système quantique est
objectivement indéterminé. La règle de Born
spécifie la probabilité de processus
fondamentalement aléatoires.

Insatisfaits de ce point de vue, Einstein
et d'autres se sont rapidement
demandé si, comme en physique statistique
classique, les probabilités ne tirent pas
leur origine d'une méconnaissance de l'état
exact du système quantique. On appelle
\emph{paramètres cachés} d'hypothétiques
éléments d'information qui s'ajouteraient
au vecteur d'état et dont la connaissance
permettrait de mieux prédire les résultats
de mesure. On peut montrer que de sérieuses
restrictions contraignent les paramètres
cachés.\footnote{Bell (1964); Kochen et Specker
(1967).} Certaines approches évitent néanmoins
ces restrictions, en particulier celle qui a
d'abord été proposée par de Broglie et, de
façon plus complète, par David
Bohm.\footnote{De Broglie (1927); Bohm (1952).
Bunge s'est intéressé à cette approche, mais
il s'en est distancié lorsque des expériences
(dont nous parlerons à la section~7) ont
fermé la porte aux paramètres cachés locaux.}

De Broglie et Bohm considèrent un ensemble
de $N$ particules (que nous supposons ici,
pour plus de simplicité, sans spin). Ils
postulent que les positions des particules
constituent des paramètres cachés. Cela
signifie que chaque particule possède,
à tout instant, une position bien définie,
qu'il nous est cependant impossible de
connaître exactement. Soit
$\psi(\mathbf{r}_1, \ldots, \mathbf{r}_N, t)$
la fonction d'onde des particules. Cette
fonction est complexe et peut s'écrire comme
\begin{equation}
\psi(\mathbf{r}_1, \ldots, \mathbf{r}_N, t)
= R(\mathbf{r}_1, \ldots, \mathbf{r}_N, t)
\exp\{ i S(\mathbf{r}_1, \ldots, \mathbf{r}_N, t) \} ,
\label{bohm}
\end{equation}
où $R$ et $S$ sont deux fonctions réelles.
La théorie se base sur les postulats
suivants:\footnote{Strictement parlant,
les postulats se rapportent à la fonction
d'onde et à l'ensemble des particules de
l'univers. Les propriétés de systèmes plus
restreints en découlent et, dans les cas qui
nous intéressent, satisfont aux deux postulats
ci-dessus.}
\begin{enumerate}
\item À tout instant~$t$, les particules ont
des positions~$\mathbf{r}_i$ bien définies,
dont on ne connaît que la densité de probabilité,
égale~à
\begin{equation}
|\psi(\mathbf{r}_1, \ldots, \mathbf{r}_N, t)|^2
= R^2(\mathbf{r}_1, \ldots, \mathbf{r}_N, t) .
\end{equation}
\item La particule~$i$ suit une trajectoire
déterministe régie par l'équation
\begin{equation}
\mathbf{v}_i (\mathbf{r}_1, \ldots, \mathbf{r}_N, t)
= \frac{\hbar}{m_i} \mbox{\boldmath $\nabla$}_i
S(\mathbf{r}_1, \ldots, \mathbf{r}_N, t) .
\label{vi}
\end{equation}
Ici, $\mathbf{v}_i$ est la vitesse de la
particule~$i$, $m_i$ sa masse et $\hbar$ est
la constante de Planck réduite.
\end{enumerate}
Le second postulat signifie, en quelque sorte,
que la fonction d'onde guide la particule
(d'où le nom \emph{d'onde pilote}). On peut
montrer que la densité de probabilité~$R^2$
évolue, par le jeu des trajectoires, de la
même manière que par l'équation de Schrödinger.
Cela implique que les prédictions statistiques
de la théorie de Bohm et de Broglie coïncident
exactement avec celles de la théorie
quantique.\footnote{Les simulations numériques
de Philippidis \emph{et al.} (1979) montrent,
de façon saisissante, que la figure d'interférence
produite par les fentes de Young s'explique à
partir de trajectoires bohmiennes.}

Voyons comment la théorie de l'onde pilote
résout le problème de la mesure. Au début du
processus, les coordonnées spatiales du système
quantique et les coordonnées spatiales des
particules constituant l'appareil ont des
valeurs bien définies. Ces valeurs sont
toutefois mal connues, et leurs densités de
probabilité sont données par les fonctions
d'onde associées aux vecteurs~$|\psi\rangle$
et~$|\alpha_0\rangle$ de l'éq.~(\ref{vn0}).
Le processus de mesure transforme le vecteur
d'état du système global selon l'éq.~(\ref{vn1}),
tandis que les trajectoires des particules
satisfont à l'éq.~(\ref{vi}). Les valeurs
finales des positions sont déterminées par
les valeurs initiales. Au terme de la mesure,
les particules de l'indicateur se trouveront
dans des positions compatibles avec un seul
des vecteurs~$|\alpha_i\rangle$. L'indicateur
marquera donc une seule valeur. Et bien que le
vecteur d'état du système global soit toujours
donné par tous les termes du membre de droite
de~(\ref{vn1}), l'évolution ultérieure du
système est à toutes fins pratiques régie par
un seul de ceux-ci (celui qui correspond à la
position de l'indicateur).

Que nous dit la théorie de l'onde pilote sur
ce qui est réel? Tous les partisans de la
théorie s'entendent sur la réalité des
particules et sur le fait que celles-ci ont,
à tout instant, des positions et des vitesses
(ainsi que, bien sûr, des masses et des charges
électriques) bien définies.\footnote{Cela ne
contredit pas le principe d'incertitude de
Heisenberg, qui se traduit par l'impossibilité
de \emph{contrôler} à la fois la position et
la vitesse d'une particule.} Par contre, les
autres grandeurs physiques (comme l'énergie,
le moment cinétique, le spin) n'ont pas, en
général, de valeurs bien définies. La mesure
de ces grandeurs ne révèle pas une valeur
préexistante, mais entraîne une réponse de
l'appareil qui dépend de l'état initial du
système et du processus d'interaction.

Si la réalité des particules ne fait aucun
doute dans la théorie de l'onde pilote,
celle de la fonction d'onde est plus
problématique. Certes la fonction d'onde
régit le mouvement des particules mais,
contrairement au champ électrique par
exemple, elle dépend non pas des coordonnées
de l'espace géométrique, mais des coordonnées
de toutes les particules. Dans ce contexte,
certains préfèrent lui attribuer un statut
nomologique plutôt qu'une réalité ontologique.
%
\section{Les mondes multiples}
Dans la théorie de l'onde pilote, le vecteur
d'état évolue toujours par l'effet d'un
opérateur unitaire. Il ne subit d'aucune
manière l'effondrement de von Neumann. Ainsi
en est-il de la théorie des {\og}états
relatifs{\fg} proposée par Hugh Everett
III,\footnote{Everett (1957).} appelée
également la \emph{théorie des mondes
multiples}.

La motivation d'Everett est venue de son
insatisfaction à l'égard de l'interprétation
de Copenhague qui, nous l'avons vu, recourt
de manière essentielle à un appareil classique
ou à un observateur situé hors du cadre de la
théorie quantique. Comment, dans ce contexte,
appliquer la théorie à l'univers en entier,
à l'extérieur duquel ne se trouvent, par
définition, ni appareil ni observateur?

Lors du processus de mesure, le vecteur
d'état du système quantique et de l'appareil
évolue selon l'éq.~(\ref{vn1}). Pour rendre
compte du fait qu'on observe un seul résultat,
von Neumann a postulé que la superposition
laisse place à un seul de ses termes.
Everett, au contraire, postule que tous les
résultats coexistent. Autrement dit, aucune
position de l'indicateur représentée
dans~(\ref{vn1}) n'est plus réelle ou moins
réelle que les autres. Et si l'indicateur est
lu par un observateur humain, aucune valeur
n'est privilégiée.

Everett montre que son hypothèse rend compte
de l'observation selon laquelle deux mesures
successives d'une grandeur physique donnent
en général le même résultat, ainsi que de
l'accord intersubjectif entre observateurs.
Il prétend également pouvoir obtenir
naturellement la règle de Born, ce sur quoi
tous les chercheurs ne s'entendent
pas.\footnote{Saunders \emph{et al.} (2010).}
Du point de vue de la réalité, cependant,
la principale question à laquelle l'approche
d'Everett fait face consiste à préciser la
nature de la multiplicité.\footnote{Marchildon
(2015b).} Pour ce faire, trois avenues ont été
proposées (chacune se subdivisant d'ailleurs
en sous-avenues). 
\begin{enumerate}
\item Les mondes multiples --- Au moment de
la mesure, le monde se scinde en de multiples
copies. Chaque copie correspond à une valeur
marquée par l'indicateur. À toutes fins
pratiques, les différentes copies évoluent
par la suite de manière indépendante. Il
semble qu'Everett lui-même ait privilégié
cette avenue, bien qu'il ne l'ait pas
spécifiquement affirmé dans ses travaux
publiés.
\item Les consciences multiples --- L'appareil,
et même le cerveau d'un observateur, se trouvent
au terme de la mesure dans un état superposé.
À chaque terme de la superposition correspond
toutefois un état de conscience, et ceux-ci
sont bien définis.
\item Les secteurs décohérents de la fonction
d'onde --- La réalité est associée à des
structures raisonnablement stables que l'on
peut identifier dans la fonction d'onde
universelle.
\end{enumerate}

Chacune de ces avenues présente une
conception de la réalité bien particulière.
La première est sans doute la plus
contre-intuitive, puisqu'on n'a aucunement
l'impression ou la sensation de se dédoubler
lors de chaque mesure quantique. À cela,
Everett répond que son approche explique
qu'on ne sente rien, un peu comme l'inertie
galiléenne explique qu'on ne se sent pas
emporté à grande vitesse par la rotation de
la Terre. La seconde avenue ne fait rien de
moins que restituer le dualisme esprit-matière.
Quant à la troisième, la plus populaire de nos
jours, elle implique que des structures qui
coexistent dans le même espace-temps (comme
le chat mort et le chat vivant) n'ont aucune
action l'une sur l'autre. D'une certaine
manière chacune est, pour ainsi dire, un
fantôme pour l'autre. Les partisans de cette
avenue doivent expliquer comment elle peut
être cohérente avec une théorie d'interactions
des constituants fondamentaux.
%
\section{Les corrélations à distance}
Les différentes interprétations de la théorie
quantique visent principalement à résoudre
le problème de la mesure, qui ne se présente
pas comme tel en théorie classique. La théorie
quantique se distingue également par
l'existence de corrélations entre systèmes
plus marquées que ce qu'on trouve en théorie
classique.

Pour illustrer la nature des corrélations
quantiques, le plus simple consiste à
examiner des particules de spin~1/2, comme des
électrons.\footnote{Le spin est proportionnel
au moment magnétique que nous avons introduit
à la section~2.} Le spin est un ensemble de
grandeurs physiques~$S_{\mathbf{n}}$, chacune
correspondant à un vecteur unitaire~$\mathbf{n}$
(ou, ce qui revient au même, à une direction
dans l'espace). Pour les électrons, chaque
grandeur~$S_{\mathbf{n}}$ ne peut prendre que
deux valeurs qui (en unités de $\hbar/2$)
sont égales à~$+1$ ou~$-1$.

On peut définir, dans l'espace d'états d'un
électron, des vecteurs $|+;\mathbf{n}\rangle$
et $|-;\mathbf{n}\rangle$ pour chaque
direction~$\mathbf{n}$. Si, par exemple,
l'électron se trouve dans l'état
$|+;\mathbf{n}\rangle$, la mesure
de~$S_{\mathbf{n}}$ donnera nécessairement
la valeur~$+1$.\footnote{Ici comme ailleurs
dans cette section, on parle de mesures
idéales. Les mesures concrètes ne sont jamais
parfaitement exactes.} Mais alors, on ne peut
prédire avec certitude le résultat de la
mesure d'aucune grandeur~$S_{\mathbf{n}'}$,
où~$\mathbf{n}'$ est une direction inclinée
par rapport à~$\mathbf{n}$.

Considérons maintenant un système quantique
constitué de deux électrons. Le vecteur
\begin{equation}
\frac{1}{\sqrt{2}} \left(
|+;\mathbf{n}\rangle |-;\mathbf{n}\rangle
- |-;\mathbf{n}\rangle|+;\mathbf{n}\rangle \right)
\label{sing}
\end{equation}
(où le premier terme d'un produit se rapporte
au premier électron, et le second terme au
second électron) appartient à l'espace d'états
de ce système. À cause du facteur de gauche,
sa norme vaut~1. Ce vecteur représente ce
qu'on appelle \emph{l'état singulet} du système
de deux électrons. Il possède deux propriétés
particulièrement importantes:
\begin{enumerate}
\item Les résultats de la mesure des
grandeurs~$S_{\mathbf{n}}$ des deux électrons
sont parfaitement anticorrélés. Cela signifie
qu'on obtient ou bien la valeur~$+1$ pour le
premier électron et la valeur~$-1$ pour le
second, ou bien l'inverse.\footnote{La façon
la plus simple de comprendre cette propriété
consiste à se placer dans le cadre de
l'effondrement du vecteur d'état. Si, par
exemple, la mesure de la
grandeur~$S_{\mathbf{n}}$ du premier
électron donne la valeur~$+1$, alors le
vecteur~(\ref{sing}) s'effondre sur le premier
terme à l'intérieur des parenthèses. La mesure
de la grandeur~$S_{\mathbf{n}}$ du second
électron donnera alors nécessairement la
valeur~$-1$.} 
\item Le vecteur~(\ref{sing}) est invariant
dans les rotations, c'est-à-dire qu'il ne dépend
pas de la direction~$\mathbf{n}$.\footnote{On
trouve la preuve de cet énoncé dans les
ouvrages de mécanique quantique.} Cela implique
que l'anticorrélation indiquée ci-dessus est
valable quelle que soit la grandeur~$S_{\mathbf{n}'}$
mesurée sur les deux électrons. 
\end{enumerate}

Nous allons supposer que l'anticorrélation que
nous avons indiquée se maintient quelle que soit
la distance entre les deux électrons et quel que
soit l'intervalle de temps séparant les deux
mesures.\footnote{Concrètement, les
expériences sont le plus souvent réalisées
avec des photons, qu'on a réussi à séparer
de plus de 1\,000\,km.} Si tel est le cas,
il ne semble y avoir que cinq explications
logiquement possibles de cette anticorrélation.
\begin{enumerate}
\item Le résultat de la mesure de n'importe
quelle grandeur~$S_{\mathbf{n}}$ est déterminé
dès la préparation de l'état singulet.
\item La mesure du spin d'un électron a un
effet instantané sur le spin de l'autre.
\item L'univers obéit à un déterminisme
intégral. Le libre choix de la direction suivant
laquelle le spin est mesuré n'est qu'une
illusion.
\item La mesure du spin agit rétroactivement
(par causalité inversée) sur la préparation
de l'état singulet.
\item La corrélation est une instance d'un
résultat extrêmement improbable, comme celui
d'obtenir, en lançant un dé non pipé, la
valeur~4 mille fois de suite.
\end{enumerate}
On pourrait penser que l'hypothèse de mondes
multiples offrirait une sixième explication
possible. Tel n'est pas le cas, cependant,
puisque les corrélations se présentent
intégralement dans chacun des mondes.

Pour autant que le présent auteur le sache,
personne ne soutient la cinquième explication.
La quatrième est défendue par John Cramer dans
ce qu'il appelle \emph{l'interprétation
transactionnelle} de la théorie
quantique.\footnote{Cramer (1986).} L'hypothèse
de causalité inversée est délicate, et ses
défenseurs doivent s'assurer qu'elle ne
donne pas lieu à des boucles causales
insidieuses.

La troisième explication paraît, à
première vue, tout à fait contraire à
l'intuition. Comment les lois de la nature
conspireraient-elles pour contraindre, à
leur insu, deux expérimentateurs à choisir
précisément les axes de mesure qui
donneraient les résultats voulus? Cette
explication est néanmoins défendue par nul
autre que Gerard 't Hooft.\footnote{'t
Hooft (2005).}

La première explication est, sans doute,
celle à laquelle la plupart songeraient
d'abord. Et en effet Einstein, Boris Podolsky
et Nathan Rosen l'ont essentiellement
proposée lorsqu'ils ont mis en lumière des
corrélations à distance analogues à celles
du spin.\footnote{Einstein \emph{et al.}
(1935).} Dans cette explication, le résultat
de la mesure du spin d'un électron selon tel
ou tel axe serait entièrement déterminé par
des paramètres cachés locaux, fixés dès la
préparation de l'état singulet et indépendants
de ce qui se passe au voisinage de l'autre
électron. Comme l'ont remarqué Einstein et ses
collaborateurs, le vecteur d'état ne
constituerait alors qu'une spécification
incomplète de l'état du système.

Malgré son attrait, la première explication
a été éliminée à la suite des travaux de
John Bell.\footnote{Bell (1964).} Celui-ci a
prouvé que l'hypothèse de paramètres cachés
locaux implique de sérieuses contraintes sur
les corrélations des mesures du spin des
électrons suivant différentes directions.
L'expérience a montré, hors de tout doute
raisonnable, que ces contraintes sont violées,
et donc qu'on ne peut soutenir la première
explication.\footnote{Dans certaines
expériences (Aspect \emph{et al.}, 1982), le
choix des directions suivant lesquelles les
spins des deux particules sont mesurés est
effectué si peu de temps avant la mesure que
même la lumière ne pourrait pas transmettre
cette information d'une particule à l'autre
en temps opportun.}

Ainsi la seconde explication, celle d'un
effet instantané de la mesure du spin d'un
électron sur le spin de l'autre, semble pour
plusieurs chercheurs inévitable. Cette
explication s'inscrit particulièrement bien
dans la théorie de l'onde pilote. En effet,
revenons à l'éq.~(\ref{vi}). On voit que
la vitesse d'une particule dépend des
positions de toutes les autres. Ce qui se
passe au voisinage de la particule~$i$ peut
donc influencer instantanément la
particule~$j$. La seconde explication trouve
également sa place dans la théorie de~GRW.
L'interprétation de Copenhague et la théorie
des mondes multiples l'incorporent moins
naturellement.

La seconde explication peut sembler,
à première vue, incompatible avec la
prohibition de signaux plus rapides que
la lumière. Le problème est subtil, mais
signalons simplement que dans le cadre de
la théorie quantique (et dans toutes ses
interprétations), l'expérimentateur n'a
qu'un contrôle limité sur la préparation
d'un état. Il ne peut pas, en particulier,
ajuster les paramètres cachés à sa guise.
On peut montrer que de cela découle
l'impossibilité de transmettre, au moyen
des corrélations à distance, quelque
information que ce soit plus vite que la
lumière. Cette observation suffit à réfuter
toute tentative de justifier, au moyen des
corrélations quantiques, de
pseudo-phénomènes comme la télépathie
et la télékinésie.
%
\section{Conclusion}
Les fondateurs de la théorie quantique se
sont rendu compte dès l'origine que la théorie
conduit à une révision importante de la
conception classique de la réalité.
L'interprétation de Copenhague, longtemps
dominante, a offert une conception instrumentaliste.
Aujourd'hui, il est possible de revenir à un
point de vue plus résolument réaliste. Il reste
que la théorie est compatible avec différentes
visions de la réalité, dont le caractère non
local semble toutefois difficilement contournable.
\section*{Remerciements}
Je remercie Jean-René Roy de ses commentaires
sur une première version du manuscrit. Je suis
reconnaissant au Conseil national de recherches
en sciences naturelles et en génie du Canada
pour son soutien pendant de nombreuses années.
%
\section*{Références}
Allori, V., Goldstein, S., Tumulka, R.
et Zanghì, N. (2008),
{\og}On the common structure of Bohmian
mechanics and the Ghirardi-Rimini-Weber
theory{\fg},
\textit{British Journal for the
Philosophy of Science}
\textbf{59}, 353--389.

\bibspace
Aspect, A., Dalibard, J. et Roger, G (1982),
{\og}Experimental test of Bell's inequalities
using time-varying analyzers{\fg},
\textit{Physical Review Letters}
\textbf{49}, 1804--1807.

\bibspace
Bell, J. S. (1964),
{\og}On the Einstein-Podolsky-Rosen paradox{\fg},
\textit{Physics} \textbf{1}, 195--200.

\bibspace
Bohm, D. (1952),
{\og}A suggested interpretation of the quantum
theory in terms of `hidden' variables, I et II{\fg},
\textit{Physical Review} \textbf{85}, 166--193.

\bibspace
Bunge, M. (1967),
\textit{Foundations of Physics}
(Springer).

\bibspace
Bunge, M. (1985),
\textit{Treatise on Basic Philosophy}, Vol.~7,
\textit{Epistemology and Methodology III:
Philosophy of Science and Technology}, Part~I,
\textit{Formal and Physical Sciences} (Reidel).

\bibspace
Cramer, J. G. (1986),
{\og}The transactional interpretation of
quantum me\-chan\-ics{\fg},
\textit{Reviews of Modern Physics}
\textbf{58}, 647--687.

\bibspace
De Broglie, L. (1927),
{\og}La mécanique ondulatoire et la structure
atomique de la matière et du rayonnement{\fg},
\textit{Le journal de physique et le radium}
\textbf{8}, 225--241.

\bibspace
Einstein, A., Podolsky, B. et Rosen, N. (1935),
{\og}Can quantum-mechanical description
of physical reality be considered complete?{\fg},
\textit{Physical Review} \textbf{47}, 777--780.

\bibspace
Everett, H. (1957),
{\og}`Relative state' formulation of
quantum mechanics{\fg},
\textit{Reviews of Modern Physics}
\textbf{29}, 454--462.

\bibspace
Fuchs, C. A., Mermin, N. D. et Schack, R. (2014),
{\og}An introduction to QBism with an application
to the locality of quantum mechanics{\fg},
\textit{American Journal of Physics}
\textbf{82}, 749--754.

\bibspace
Ghirardi, G. C., Rimini, A. et Weber, T. (1986),
{\og}Unified dynamics for microscopic and
macroscopic systems{\fg},
\textit{Physical Review D}
\textbf{34}, 470--491.

\bibspace
Ghirardi, G. C., Pearle, P. et Rimini, A. (1990),
{\og}Markov processes in Hilbert space and
continuous spontaneous localization of systems
of identical particles{\fg},
\textit{Physical Review A}
\textbf{42}, 78--89.

\bibspace
Jammer, M. (1974),
\textit{The Philosophy of Quantum Mechanics}
(Wiley).

\bibspace
Kochen, S. et Specker, E. P. (1967),
{\og}The problem of hidden variables in
quantum mechanics{\fg},
\textit{Journal of Mathematics and Mechanics}
\textbf{17}, 59--87.

\bibspace
London, F. et Bauer, A. (1939),
{\og}La théorie de l'observation en mécanique
quantique{\fg},
\textit{Actualités scientifiques et industrielles},
N\xp{o} 775 (Hermann).

\bibspace
Marchildon, L. (2000),
\textit{Mécanique quantique}
(De Boeck).

\bibspace
Marchildon, L. (2015a),
{\og}Why I am not a QBist{\fg},
\textit{Foundations of Physics}
\textbf{45}, 754--761. 

\bibspace
Marchildon, L. (2015b),
{\og}Multiplicity in Everett's interpretation
of quantum mechanics{\fg},
\textit{Studies in History and Philosophy
of Modern Physics} \textbf{52}, 274--284.

\bibspace
Philippidis, C., Dewdney, C. et Hiley, B. J. (1979),
{\og}Quantum interference and the quantum
potential{\fg},
\textit{Nuovo Cimento}
\textbf{52B}, 15--28.

\bibspace
Popper, K. R. (1982),
\textit{La théorie quantique et le schisme
en physique}
(traduction française, Hermann, 1996).

\bibspace
Saunders, S., Barrett, J., Kent, A. et
Wallace, D., éds. (2010),
\textit{Many Worlds? Everett,
Quantum Theory, \&\ Reality}
(Oxford University Press).

\bibspace
Schrödinger, E. (1935),
{\og}The present situation in quantum
mechanics{\fg},
traduction anglaise dans
\textit{Proceedings of the American
Philosophical Society}
\textbf{124}, 323--338 (1980).

\bibspace
't Hooft, G. (2005),
{\og}Determinism beneath quantum mechanics{\fg},
dans \textit{Quo Vadis Quantum Mechanics?},
A. C. Elitzur, S. Dolev et N. Kolenda, éds.
(Springer), 99--111.

\bibspace
Ulfbeck, O. et Bohr, A. (2001),
{\og}Genuine fortuitousness. Where did
that click come from?{\fg},
\textit{Foundations of Physics}
\textbf{31}, 757--774.

\bibspace
Von Neumann, J. (1932),
\textit{Les fondements mathématiques de la
mécanique quantique}
(traduction française, Félix Alcan, 1946).

\bibspace
Wigner, E. P. (1961),
{\og}Remarks on the mind-body question{\fg},
dans \textit{The Scientist Speculates},
I.~J.~good, éd.\
(William Heinemann), 284--302.
\end{document}